\documentstyle[prl,aps,psfig]{revtex}
\begin{document}
\begin{center}
{\Large\bf d=3 random field behavior near percolation}

\vspace{0.25in}

{F. C. Montenegro}\\
{\small Departamento de Fisica, Universidade Federal de Pernambuco,
50670-901 Recife PE, Brasil}\\
{D. P. Belanger and Z. Slani\v{c}}\\
{\small Department of Physics, University of California,
Santa Cruz, California 95064}\\
{J. A. Fernandez-Baca}\\
{\small Solid State Division, Oak Ridge National Laboratory,
Oak Ridge, Tennessee 37831-6393}
\end{center}
\vspace{0.5in}
\begin{abstract}
The highly diluted antiferromagnet $Mn_{0.35}Zn_{0.65}F_2$ has been
investigated by neutron scattering for $H>0$.
A low-temperature ($T<11$~K), low-field ($H<1T$) pseudophase
transition boundary separates a partially antiferromagnetically
ordered phase from the paramagnetic one.
For $1<H<7$~T at low temperatures, a region of antiferromagnetic
order is field induced but is not enclosed within
a transition boundary.

\end{abstract}
\vspace{0.4in}

Diluted uniaxial antiferromagnets (AF) in applied fields are ideal
random-field Ising model (RFIM) systems\cite{b98,fa79,c84}.
$Mn_{0.35}Zn_{0.65}F_2$ is a three dimensional ($d=3$)
system with a magnetic concentration $x$ close to the percolation
threshold\cite{e80}, $x_p=0.25$.  Its behavior may
be contrasted with $Fe_xZn_{1-x}F_2$, which differs magnetically from
$Mn_xZn_{1-x}F_2$ only in the nature and strength of the
anisotropy, which is dipolar in the latter\cite{nld69} and is dominated
by a much larger single-ion anisotropy in
the former\cite{hrg70}.  Samples of $Fe_xZn_{1-x}F_2$ have been investigated
with $x=0.25$, $0.27$ and $0.31$.  The first two do
not order even with $H=0$ and show spin-glass-like
behavior\cite{mrc88,by93,jdnb97,si98,jnm99}
at all $H$.  The $x=0.31$ sample shows AF long-range order (LRO) 
at low $H$ and spin-glass-like behavior at higher
fields\cite{mlcr90,mkjhb91,bmmkj91}.
In $Mn_{x}Zn_{1-x}F_{2}$, ac susceptibility measurements indicate a
spin-glass-like clustering at low temperatures for samples with
$0.2<x<0.35$\cite{bp81,mrjmm95}.  We present evidence suggesting that frozen
spin-glass-like clusters also affect the stability of the
AF long-range order in $Mn_{0.35}Zn_{0.65}F_{2}$.

The neutron scattering experiments were performed at the Oak Ridge National
Laboratory High Flux Isotope Reactor using a two-axis configuration
with a monochromated $14.7$~meV beam.  Further experimental
details have been previously reported\cite{msbf99}.
For simplicity, the transverse line shapes used in fits of the data
for $H>0$ are of the mean-field RFIM form\cite{b98}
\begin{equation}
S(q)=\frac{A}{q^2+\kappa^2}+\frac{B}{(q^2+\kappa^2)^2} \quad ,
\end{equation}
where $\kappa$ is the inverse correlation length for fluctuations.
We do not fit data in the Bragg scattering region $|q|<0.008$ reciprocal
lattice units (rlu).

Figure 1 shows AF (100) transverse scans obtained by
heating with $H>0$ after cooling in zero field (ZFC) and the fits to
Eq.\ 1 for $H=0.35$~T.  A striking feature of these scans is
the relatively small change in the line shape with temperature when
compared with the $H=0$ scans reported previously\cite{msbf99},
particularly for $T>T_c(H)$, where $T_c(H)$ is a transition-like
temperature.  The fits to the data are essentially identical
whether $B$ is fixed to zero or allowed to vary, indicating that
the squared-Lorentzian term in Eq.\ 1 is unimportant.  The results
for $\kappa$, which reflect the unusual line shape behavior, are shown
in Fig.\ 2.  The values for $H=0.35$~T drop well below the
values for $H=0$ which are concentration-gradient limited,
indicating that Eq.\ 1 does not describe the line shapes very well
for $H>0$.  The clear minimum in $\kappa$ vs.\ $T$ and Bragg scattering
intensities vs.\ $T$ (not shown) do indicate that the system is trying to
undergo a transition, but it is questionable that one successfully occurs.
For $H=0.75$~T, when $B$ is fixed to zero, the values of $\kappa$ are
artificially much smaller than the instrumental resolution.
The results when $B$ is allowed to vary are those
shown in Fig.\ 2.  The minimum in $\kappa$ vs.\ $T$
is extremely shallow although a transition-like region is indicated by
a peak in the scattering at $|q|=0.008$~rlu vs.\ $T$ as well
as the rapid decline in the Bragg scattering intensity.  Either
a transition does
not actually take place or very few spins are involved.
From temperatures at which the maxima in the off-Bragg scattering at
$|q|=0.008$~rlu and the maxima in the temperature derivative of the
Bragg scattering with respect to temperature occur,
the AF pseudophase boundary is determined.  The boundary is shown by the points
with horizontal error bars in the inset of Fig.\ 3.  The points with vertical
error bars indicate the magnetic field at which an
intensity change is observed in the Bragg scattering and
at which the peak of the off-Bragg scattering occurs as the field is
changed while $T$ is kept constant.  The transition region again appears
quite broad.  An example of this at $T=5K$ is shown in Fig.\ 3
which also indicates substantial hysteresis when comparing
ZFC data and data obtained upon cooling in the field (FC) for both $q=0$ and
$|q|=0.008$~rlu and is consistent with earlier magnetization
measurements\cite{mrjmm95}.

For $2<H<7$~T a region of AF LRO is indicated
by the $q=0$ intensities in Fig.\ 4.
The order is most intense at
$H \approx 5$~T.  There is no peak in the off-Bragg ($|q|=0.008$~rlu)
scattering for $T>5$~K, as exemplified for $H=5$~T in the inset of Fig.\ 4.
This indicates that there is no phase transition boundary associated with
this AF order and that the AF order is most likely field induced.
Within a cluster the sublattice with an excess of spins
will preferentially order along the field direction.  The other
sublattice will then order in the opposite direction as a result of the
exchange interaction.
At fields above $5$~T, the AF order weakens.
This contrasts the behavior observed\cite{mjmmr92} in $Mn_{0.5}Zn_{0.5}F_2$,
where a spin-flop phase with a clear transition boundary
occurs above the AF one.

Previous magnetization measurements were made of the pseudotransition
boundary\cite{mrjmm95}.  The inset in Fig.\ 3 shows the peak positions
of $d(MT)/dT$ for $H<0.5T$ as the starred data points.  The widths of these
peaks are very large.  For example, the half width at half maximum at $H=0.35T$
is more than $1$~K.  These widths are much larger
compared to measurements at larger $x$\cite{mjmmr92,rkj88}.
Note that the positions do not coincide
with the boundary determined with neutrons and even the curvature
is different.  For the neutron boundary, $T-T_c(H) \sim H^{2/\phi}$ with
$\phi = 1.4$ as in RFIM systems further from $x_p$\cite{b98}.  In contrast,
the magnetization data are fit with $\phi=3.4$\cite{mrjmm95}, the typical
spin-glass exponent. Above $H=0.5$~T, it was not possible to reliably determine
a peak position.  These results reinforce the scenario in which the
boundaries do not represent a true transition to AF LRO,
but rather an AF transition that is strongly interfered with
by spin-glass-like frozen clusters.  The transition at $T=11$~K for $H=0$,
in contrast, seems much more normal,
though spin-glass-like behavior is evident\cite{msbf99} below $T=7$~K.
Slow relaxation seems evident in both the neutron scattering and magnetization
for $H>0$, probably a result of clustering induced by the alignment
of domains by the field.  At low $H$, hysteresis is observed well above
$T_c(H)$ and clustering is indicated by a strong deviation from
Curie-Weiss behavior below $T=22$~K\cite{mrjmm95}.
The small anisotropy in the $Mn_xZn_{1-x}F_2$
system allows the field to align clusters more easily\cite{bp81} than
in $Fe_xZn_{1-x}F_2$ and this effect certainly must account for the differences
between these two systems.

This work has been supported
by DOE Grant No. DE-FG03-87ER45324
and by ORNL, which is managed by
Lockheed Martin Energy Research Corp. for the U.S. DOE
under contract number DE-AC05-96OR22464.
One of us (F.C.M.) also acknowledges the support of
CAPES, CNPq, FACEPE and FINEP (Brazilian agencies).

\newpage

\begin{figure}[t]
\centerline{\hbox{
\psfig{figure=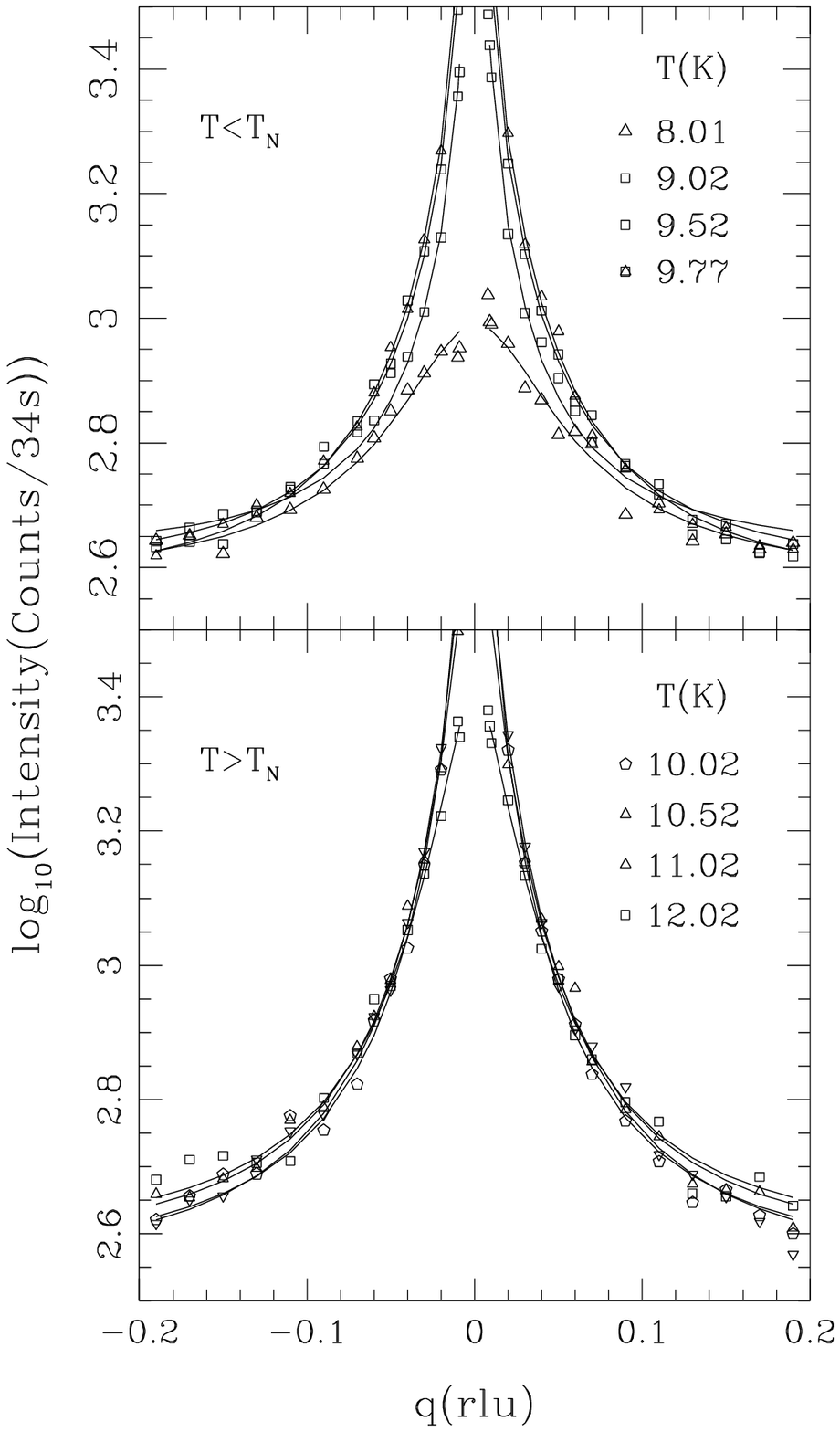,height=6.0in}
}}
\caption{The logarithm of the ZFC scattering intensity vs.\ $q$ for
various temperatures above and below $T_c(H)$ for $H=0.35$~T.
}
\end{figure}

\newpage

\begin{figure}[t]
\centerline{\hbox{
\psfig{figure=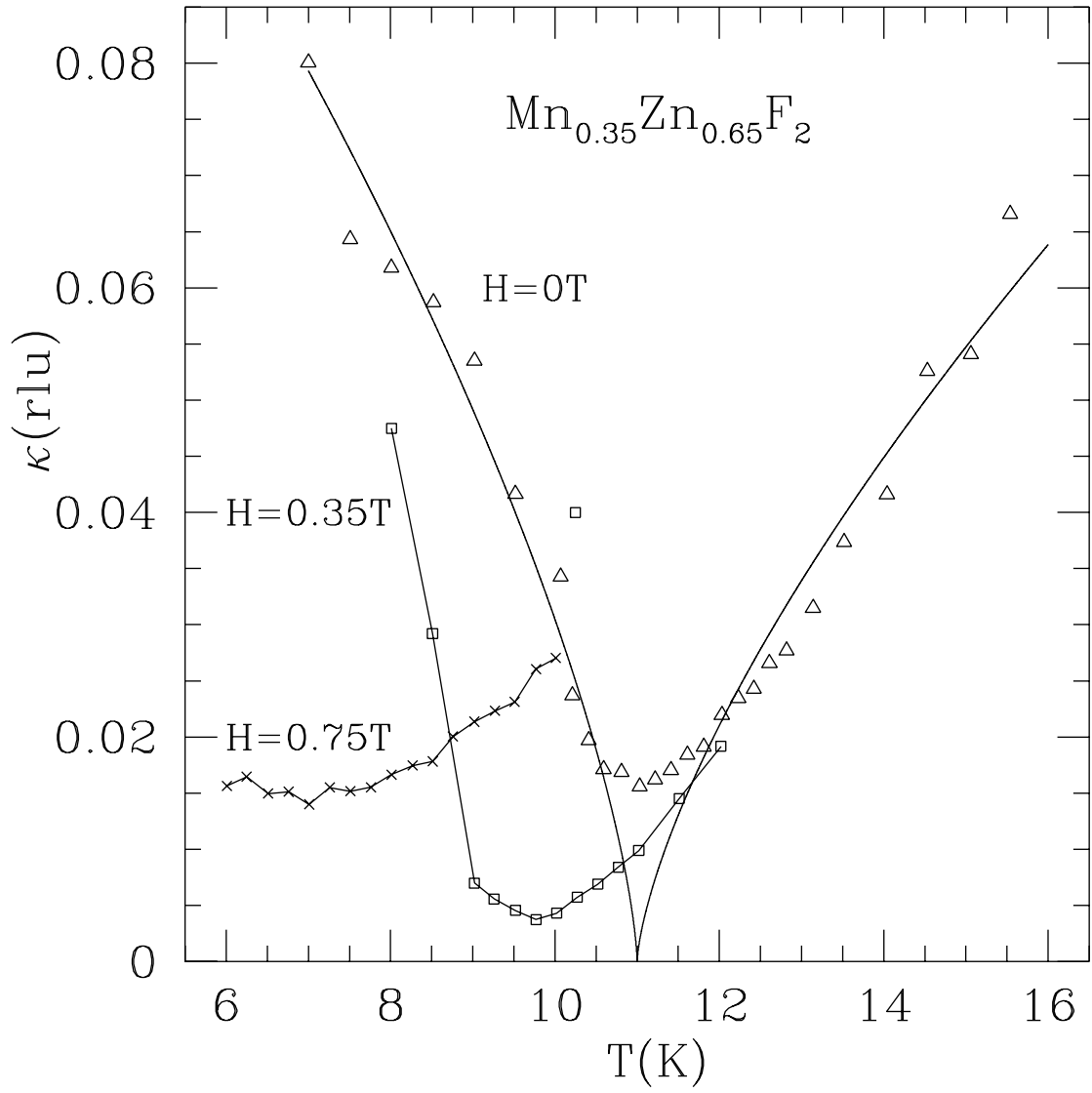,height=6.0in}
}}
\caption{The results for $\kappa$ vs.\ $T$ for $H=0$, $0.35$ and $0.75$~T
using Eq.\ 1.
}
\end{figure}

\newpage

\begin{figure}[t]
\centerline{\hbox{
\psfig{figure=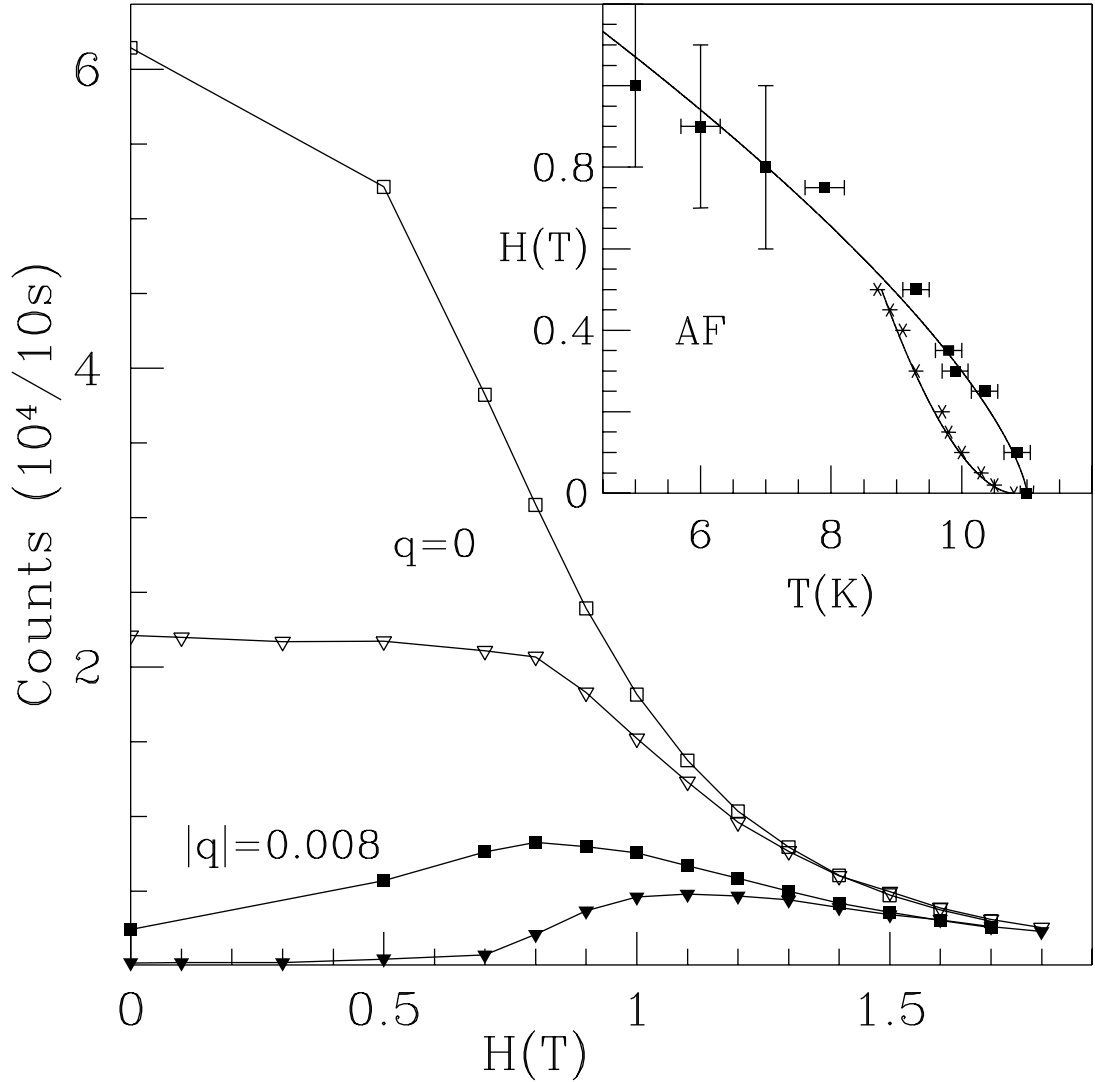,height=6.0in}
}}
\caption{The scattering intensity vs.\ $H$
at $q=0$ and $|q|=0.008$~rlu for field increasing (triangles) and
field decreasing (squares) after zero-field cooling to $T=5$~K.
The inset shows the pseudophase boundary determined from
neutron scattering (squares) and magnetization (stars) measurements.
}
\end{figure}

\newpage

\begin{figure}[t]
\centerline{\hbox{
\psfig{figure=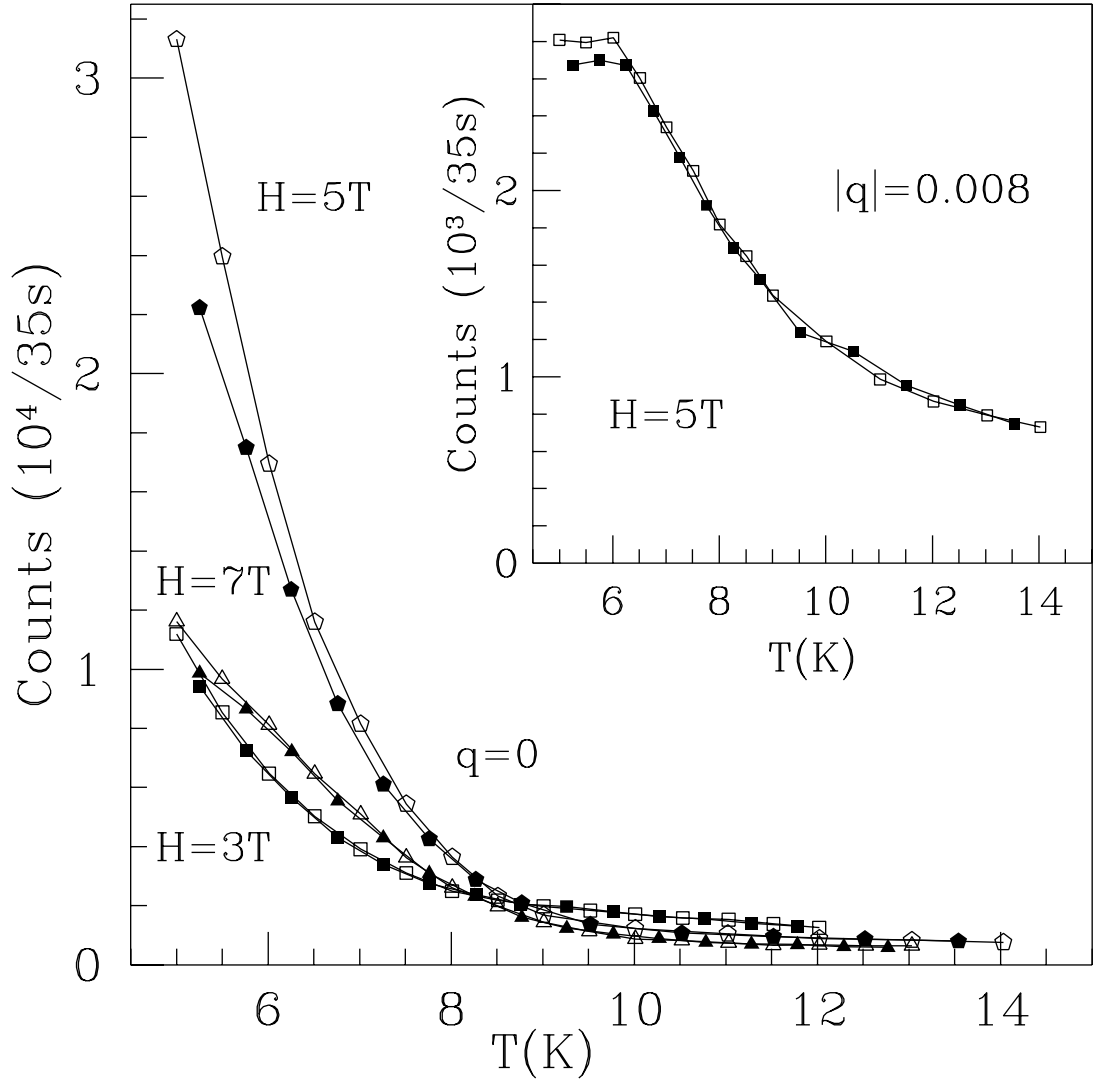,height=6.0in} 
}}
\caption{The scattering intensity vs.\ $T$ at $H=3$, $5$ and $7$~T
at $q=0$~rlu upon heating after ZFC
(open symbols) and upon FC (filled symbols).
The inset shows the scattering intensity vs.\ $T$ for $H=5$~T
at $|q|=0.008$~rlu after ZFC
(open symbols) and upon FC (filled symbols).
}
\end{figure}

\end{document}